\documentclass[
preprintnumbers]{revtex4}
\usepackage{graphicx}

\usepackage{epsbox}
\usepackage{amsmath,amssymb}
\usepackage{booktabs}
\usepackage{bm}

\begin{document}
%
%
\preprint{\vtop{
{\hbox{YITP-08-27}\vskip-0pt
                 \hbox{KANAZAWA-08-03} \vskip-0pt
}}}
\thispagestyle{empty}
\title{Charmed Scalar Resonances\footnote{Talk given at SCADRON70,  
the workshop on ``Scalar Mesons and Related Topics'', 11 -- 16 February 
2008, IST-Lisbon, Lisbon, Portugal.}
}
\author{Kunihiko Terasaki} 
\affiliation{ Yukawa Institute for Theoretical Physics, 
Kyoto University, Kyoto 606-8502, Japan,\\
Institute for Theoretical Physics, Kanazawa University, 
Kanazawa 920-1192, Japan
}
\thispagestyle{empty}



\begin{abstract}
 It is pointed out that assigning $D_{s0}^+(2317)$ to 
$\hat F_I^+\sim [cn][\bar s\bar n]_{I=1}$ is favored by experiment. It 
also is discussed why its neutral and doubly charged partners have never 
been observed in inclusive $e^+e^-$ annihilation. To search for them, 
hadronic weak decays of $B$ mesons would be better, and their production 
rates would be 
$Br(B_u^+\rightarrow D^-\hat F_I^{++})
\sim Br(B_d^0\rightarrow \bar D^0\hat F_I^{0})\sim 10^{-3}$. 
\end{abstract}

\maketitle

The charm-strange scalar meson $D_{s0}^+(2317)$ was discovered in 
inclusive $e^+e^-$ annihilation~\cite{Babar-D_s,CLEO-D_s}. Its mass and 
width are now compiled as~\cite{PDG06}, 
$m_{D_{s0}}=2317.3 \pm 0.6\,\,{\rm MeV}$, 
$\Gamma(D_{s0}^+(2317)) < 4.6\,\,{\rm MeV},\,\,{\rm CL=90}\, \%$. 
However, its signal has never been observed in the radiative 
$D_s^{*+}\gamma$ channel. Therefore, a severe constraint, 
\begin{eqnarray}
\mathcal{R}(D_{s0}^+(2317))_{\rm exp} 
= \frac{\Gamma(D_{s0}^+(2317)\rightarrow D_{s}^{*+}\gamma)}
{\Gamma(D_{s0}^+(2317)\rightarrow D_{s}^{+}\pi^0)}\Bigl|_{\rm exp} 
< 0.059,
                                            \label{eq:constraint-D_s}
\end{eqnarray}
has been provided~\cite{CLEO-D_s}. For reference, we here list another 
data on the ratio~\cite{PDG06}, 
\begin{equation}
\mathcal{R}(D_s^{*+})^{-1}_{\rm exp}  
=\frac{\Gamma(D_s^{*+}\rightarrow D_s^+\pi^0)}
{\Gamma(D_s^{*+}\rightarrow D_s^+\gamma)}\Bigl|_{\rm exp}
=0.062\pm 0.006. 
                                         \label{eq:D_s^*-constraint}
\end{equation}
Eq.~(\ref{eq:D_s^*-constraint}) implies that the isospin non-conserving 
interaction is much weaker than the electromagnetic interaction. With 
this in mind, it is learned, from Eq.~(\ref{eq:constraint-D_s}), that 
the interaction causing the decay in the denominator is much stronger 
than the electromagnetic interaction, i.e., it is the well-known strong 
interaction which conserves isospin. Therefore, $D_{s0}^+(2317)$ should 
be an isospin $I=1$ state which cannot be realized by the conventional 
$\{c\bar s\}$ but can be realized by a tetra-quark meson, 
$\hat F_I^+\sim [cn][\bar s\bar n]_{I=1}$ with $n=u,\,d$. (Notation of 
the tetra-quark mesons will be given later.) It is also learned that 
its $I=0$ partner $\hat F_0^+\sim [cn][\bar s\bar n]_{I=0}$ would be 
observed in the $D_s^{*+}\gamma$ channel. 

In addition, charm-strange scalar mesons which are degenerate with 
$D_{s0}^+(2317)$ have been observed in $B$ 
decays~\cite{BELLE-D_s,Babar-B}. In particular, some indications of 
existence of a charm-strange scalar meson have been observed in the 
radiative channel~\cite{BELLE-D_s}. It is quite different from the 
above $D_{s0}^+(2317)$. Therefore, the charm-strange scalar mesons 
observed in $B$ decays are denoted by 
$\tilde D_{s0}^+(2317)[{observed\,\, channel}]$ 
to distinguish it from the above $D_{s0}^+(2317)$, although 
$\tilde D_{s0}^+(2317)[D_s^+\pi^0]$ will be identified with 
$D_{s0}^+(2317)$ later. 

Regarding charmed non-strange scalar mesons, a broad enhancement ($D_0$) 
just below the tensor $D_2^*$ in $D\pi$ mass distribution has been 
observed~\cite{Belle-D_0,FOCUS}. Its mass and width are now compiled 
as~\cite{PDG06} 
\begin{equation}
m_{D_{0}}=2352 \pm 50\,\,{\rm MeV}, \quad  
\Gamma(D_{0})_{\rm exp} =261 \pm 50\,\,{\rm MeV}.         \label{eq:D_0} 
\end{equation}
To check if $D_0$ can be interpreted as the scalar 
$D_0^*\sim\{c\bar n\}$ meson, we here study two-body decays of $D_0^*$ 
(and its strange partner $D_{s0}^{*+}$) by using a hard pion (or kaon) 
technique in the infinite momentum frame~\cite{hard-pion,suppl}, which 
is an innovation of the old current algebra. (For the technical details 
in this article, see references quoted.) In this approximation, the 
amplitude is given by {\it asymptotic} matrix element(s) (matrix 
element(s) taken between single hadron states with infinite momentum) of 
the axial charge $A_\pi$ (or $A_K$). Then, we use asymptotic flavor 
symmetry (flavor symmetry of asymptotic matrix elements) which is a 
prescription to treat broken flavor symmetry~\cite{suppl}. Comparing 
$D_0^*$ and $D_{s0}^{*+}$ with the experimentally known 
$K_0^*$~\cite{PDG06} which is assigned to the scalar 
$\{n\bar s\}$~\cite{CT}, and using asymptotic $SU_{f}(4)$ 
symmetry~\cite{TM,ECT-talk}, we estimate rates for the 
$D_0^{*}\rightarrow D\pi$ and $D_{s0}^{*+}\rightarrow (DK)^+$ decays, 
where the spatial wavefunction overlap is still in the $SU_f(4)$ 
symmetry limit at this stage. Correcting its deviation (reducing the 
overlap in open-charm meson decays by $\sim 20-30$ \% when they are 
compared with those for light-meson decays~\cite{TM,ECT-talk}) from 
the $SU_f(4)$ symmetry limit, we obtain  
$\Gamma(D_0^{*}\rightarrow D\pi)\sim 40-50$  MeV and 
$\Gamma(D_{s0}^{*+}\rightarrow (DK)^+) \simeq 30-40$ MeV, 
where the mass of $D_{s0}^{*+}$ has been estimated to be 
$m_{D_{s0}^{*+}}\simeq 2.45$ GeV by using a quark counting with 
$\Delta_s\simeq m_s-m_n\simeq 100$ MeV and $m_{D_0^*}\simeq 2.35$ GeV in 
Eq.~(\ref{eq:D_0}) as the input data. Because the above rates saturate 
approximately the full widths of $D_0^*$ and $D_{s0}^{*+}$, their 
estimated widths~\cite{TM} are 
$\Gamma(D_0^*)\simeq 40 - 50\,\,{\rm MeV}$  and 
$\Gamma(D_{s0}^{*+})\sim 30 -40\,\,{\rm MeV}$.  
It should be noted that the above $\Gamma(D_0^*)$ is much smaller than 
$\Gamma(D_0)_{\rm exp}$ in Eq.~(\ref{eq:D_0}). Therefore, we expect that 
$D_0$ has a structure including $D_0^*$. 

Four-quark mesons can be classified into the following four 
groups~\cite{Jaffe},   
\begin{equation} 
\{qq\bar q\bar q\} =  
[qq][\bar q\bar q] \oplus (qq)(\bar q\bar q)  
\oplus \{(qq)[\bar q\bar q] \pm [qq](\bar q\bar q)\} 
                                                   \label{eq:4-quark} 
\end{equation} 
with $q=u,\,d,\,s$, 
where parentheses and square brackets denote symmetry and anti-symmetry  
of flavor wavefunction, respectively, under exchange of flavors between 
them. Each term on the right-hand-side (r.h.s.) of 
Eq.~(\ref{eq:4-quark}) is again classified into two classes with 
${\bf \bar 3_c}\times{\bf 3_c}$ and ${\bf 6_c}\times {\bf \bar 6_c}$  
of the color $SU_c(3)$.  However, these two states can largely mix with 
each other in the light tetra-quark mesons while, in the corresponding 
open-charm mesons, such a mixing would be much smaller, because QCD is 
non-perturbative at the energy scale of the light meson mass while 
(rather) perturbative at the energy scale of the open-charm meson mass. 
Anyway, we here concentrate on the $[qq][\bar q\bar q]$ mesons. 
The observed mass hierarchy of low lying scalar nonet mesons, 
$a_0(980)$, $f_0(980)$, $f_0(600)$~\cite{PDG06} and  
$\kappa(800)$~\cite{E791}, and the approximate degeneracy between  
$a_0(980)$ and $f_0(980)$ can be easily understood by assigning them  
to the $[qq][\bar q\bar q]$  mesons~\cite{Jaffe}.  

Extension of the above $[qq][\bar q\bar q]$ mesons to open-charm mesons 
is straightforward: $\hat F_I\sim [cn][\bar s\bar n]_{I=1}$ and 
$\hat F_0^+\sim [cn][\bar s\bar n]_{I=0}$ with $S=1$;  
$\hat D\sim [cn][\bar u\bar d]$ and $\hat D^s\sim [cs][\bar n\bar s]$   
with $I=1/2$ and $S=0$; $\hat E^0\sim [cs][\bar u\bar d]$ with $S=-1$. 
However, it should be noted that their color configuration would be very 
much different from that of the light four-quark mesons, i.e., the 
former would be dominantly ${\bf \bar{3}_c\times 3_c}$ because of the 
attractive force between two quarks (and two antiquarks)~\cite{Hori}. 
With this in mind, we assign $D_{s0}^{+}(2317)$ to 
$\hat F_I^+$~\cite{Terasaki-D_s}. The masses of the other members are 
estimated very crudely as 
$m_{\hat D}\simeq 2.22$ GeV, $m_{\hat D^s}\simeq 2.42$ GeV, 
$m_{\hat E}\simeq 2.32$ GeV,  
by using the above quark counting and taking $m_{\hat F_I^+}\simeq 2.32$ 
MeV as the input data. The observed narrow width of $D_{s0}^{+}(2317)$ 
can be understood by a small rate for its dominant decay. It can be 
understood by a small overlap of color and spin wavefunctions which can 
be seen by decomposing the four-quark state into a sum of products of 
quark-antiquark pairs~\cite{ECT-talk,HT-isospin}. To see numerically the 
narrow width, we compare the $\hat F_I^+\rightarrow D_s^+\pi^0$ decay 
with the $\hat\delta^s\rightarrow \eta\pi$, where $a_0(980)$ has been 
assigned to $\hat\delta^s\sim [ns][\bar n\bar s]_{I=1}$.  
However, we should be careful when we compare decays with different 
energy scales, in particular, the decays of open-charm scalar mesons  
with those of the light scalar mesons. Using the hard pion technique 
with the asymptotic $SU_f(4)$ symmetry~\cite{hard-pion,suppl}, 
taking~\cite{ECT-talk,HT-isospin} $|\beta_0|^2=1/12$ describing the 
overlap of color and spin wavefunctions in the open-charm meson decays 
and 
$\Gamma(\hat\delta^s\rightarrow \eta\pi)_{\rm exp}=50 - 100$ 
MeV~\cite{PDG06} 
as the input data, and correcting the $SU_f(4)$ symmetry breaking, 
we obtain~\cite{ECT-talk,HT-isospin} a sufficiently narrow width, 
\begin{equation}
\Gamma(\hat F_I^+)\simeq \Gamma(\hat F_I^+\rightarrow D_s^+\pi^0)
\sim 2-5\,\, {\rm MeV}.                          \label{eq:width-F_I}
\end{equation}
In the same way, it is seen~\cite{Terasaki-D_s} that all the scalar 
$[cq][\bar q\bar q]$ mesons are narrow. 
\begin{center}
\begin{table}[t]
\caption{
Rates for radiative decays of charm-strange mesons. Input 
data are taken from Ref.~\cite{PDG06}. 
}
\vspace{1mm}
\begin{tabular}
{c  c  c  c  c}
\hline
 Decay & pole 
& 
{${|\beta_1|^2}$}
& Decay rate 
({keV})\hspace{-0mm}
& Input Data 
\\
\hline
{${D_s^{*+}\rightarrow D_s^+\gamma}$} & $\phi$, $\psi$ 
&\hspace{-2mm}{1}\hspace{-2mm} 
& {0.4}
&{${\Gamma(\omega\rightarrow \pi^0\gamma)_{\rm exp}}$  
${=0.734\pm 0.035}$ MeV}  
\\
\hline
{${D_{s0}^{*+}\rightarrow D_s^{*+}\gamma}$} & $\phi$, $\psi$ 
&\hspace{-2mm} {1}\hspace{-2mm} 
& {${15-20}$}  
& { ${\Gamma(\chi_{c0}\rightarrow \psi\gamma)_{\rm exp}}$  
${=119\pm 15}$ keV}  
\\   
\hline  
{$\hat{{F}}_{{0}}^{{+}}  
{\rightarrow D_s^{*+}\gamma}$} & $\omega$  
&\hspace{-2mm} {${{1}/{4}}$}\hspace{-2mm}  
& {${2 - 3}$}  
& $\Gamma(\phi\rightarrow a_0(980)\gamma)_{\rm exp}$ \vspace{0mm}
\\
\cline{1-4}
& & & &  \vspace{-3mm}\\
{$\hat{{F}}_{{I}}^{{+}}  
{\rightarrow D_s^{*+}\gamma}$} & $\rho^0$  
&\hspace{-2mm} {${{1}/{4}}$}\hspace{-2mm}  
& {${20-25}$}  
& $=0.32\pm 0.03$ keV
\\
\hline
\end{tabular}
\end{table}
\end{center}

Now we study radiative decays of $D_s^{*+}$, $D_{s0}^{*+}$, $\hat F_0^+$ 
and $\hat F_I^+$ under the vector meson dominance with the broken 
$SU_f(4)$ symmetry and the overlap factor~\cite{ECT-talk,HT-isospin}  
($|\beta_1|^2=1/4$) between $\hat F_I^+$ (or $\hat F_0^+$) and two 
vector mesons, while $|\beta_1|^2=1$ in the case of the $\{c\bar s\}$ 
meson decays because their color and spin configuration is unique. The 
results are listed in {TABLE~I} in which the $SU_f(4)$ symmetry 
breaking has been corrected. The ratio of the rate for the 
$\hat F_I^+\rightarrow D_s^{*+}\gamma$ in {TABLE~I} to that for the 
$\hat F_I^+\rightarrow D_s^{+}\pi^0$ in Eq.~(\ref{eq:width-F_I}) is 
\begin{equation}
\mathcal{R}(\hat F_I^+) \simeq (4.5 - 9)\times 10^{-3},
                                                 \label{eq:F_I-ratio}
\end{equation}
which satisfies well Eq.~(\ref{eq:constraint-D_s}). Therefore, the 
experiment favors assigning $D_{s0}^+(2317)$ to $\hat F_I^+$, as 
expected intuitively. 
\begin{center} 
\begin{table}[t] 
\caption{
Rates for isospin non-conserving decays of charm-strange 
mesons. Input data are taken from Ref.~\cite{PDG06}. 
}
\vspace{1mm}  
\begin{tabular}
{c  c  c  c }
\hline
Decay  
&  
{${|\beta_0|^2}$} 
& Input Data 
&  
Decay rate 
({keV})\hspace{-0mm} 
\\
\hline
{${D_s^{*+}\rightarrow D_s^+\pi^0}$}  
& {1}  
&  
${\Gamma(\rho\rightarrow \pi\pi)_{\rm exp}}$ 
${\simeq 150}$ MeV 
& {0.025} 
\\ 
\hline 
{${D_{s0}^{*+}\rightarrow D_s^+\pi^0}$} 
& {1}
& 
{
${\Gamma(K_0^{*0}(1430)\rightarrow K^+\pi^-)_{\rm exp}}$ 
${= 182\pm 24}$ MeV 
} 
& {${0.3}$} 
\\ 
\hline 
{$\hat{{F}}_{{0}}^{{+}} 
{\rightarrow D_s^{+}\pi^0}$}  
&\hspace{-2mm} {${{1}/{12}}$}\hspace{-2mm}  
& \hspace{2.5mm}  
$\Gamma(a_0(980)\rightarrow \eta\pi)_{\rm exp}= 50-100$ MeV  
& {${0.2-0.5}$} 
\\
\hline
\end{tabular}
\end{table}
\end{center} 

As usual~\cite{Dalitz}, we assume that isospin non-conserving decays 
proceed through a tiny  $\pi^0$-$\eta$ mixing, i.e., 
$\pi^0_{\rm phys}\simeq \pi^0 + \epsilon \eta, \,\,
(\epsilon = 0.0105 \pm 0.0013)$. 
The hard pion technique with the $\pi^0$-$\eta$ mixing and the 
asymptotic $SU_f(4)$ symmetry lead~\cite{ECT-talk,HT-isospin} to the 
rates listed in {TABLE~II}, where the $SU_f(4)$ symmetry breaking has 
been corrected. They are much smaller than the rates for the radiative 
decays of the corresponding parents, as expected intuitively. From the 
rates in~{TABLE~I} and {TABLE~II}, we obtain 
$\mathcal{R}(D_s^{*+})^{-1}\simeq 0.06$ which reproduces well 
Eq.~(\ref{eq:D_s^*-constraint}). This implies that the present approach 
is sufficiently reliable. On the other hand, our results on 
$\mathcal{R}(\hat F_0^+)$ and $\mathcal{R}(D_{s0}^{*+})$ are much larger 
than the experimental upper bound. Therefore, assigning $D_{s0}^+(2317)$ 
to $\hat F_0^+$ or $D_{s0}^{*+}$  should be excluded. 

Independently of the above discussions, $\mathcal{R}(D_{s0}^+(2317))$ 
has been studied by assigning $D_{s0}^+(2317)$ to $D_{s0}^{*+}$ (or the 
chiral partner of $D_s^+$). Although some of them~\cite{mild} provided 
values of $\mathcal{R}(D_{s0}^{*+})$ smaller than unity in contrary to 
our intuitive and numerical discussions, these results are still beyond 
the experimental upper bound. The other results~\cite{large-m_s} are 
close to the upper bound of Eq.~(\ref{eq:constraint-D_s}) or satisfy 
it. However, these theories have taken a large $s$-quark mass which is 
incompatible with the heavy $c$-quark picture and a large 
$\Gamma(D_{s0}^{*+}\rightarrow D_s^+\pi^0)$ which leads to a huge 
$\Gamma(D_0^*)$ beyond $\Gamma(D_0)_{\rm exp}$ in Eq.~(\ref{eq:D_0}). 
The remaining model is a unitarized one~\cite{Faessler} in which 
$D_{s0}^+(2317)$ is assigned to a $\{DK\}_{I=0}$ molecule. In this case, 
the mechanism to cause the isospin non-conservation is more complicated 
than the usual one, and the results are strongly dependent on the values 
of parameters involved. In this model, however, the charm-strange 
axial-vector meson $D_{s1}^+(2460)$ is interpreted as a $\{D^*K\}$ 
molecule~\cite{Faessler-1}, and its ratio $\mathcal{R}(\{D^*K\})$ 
of decay rates corresponding to $\mathcal{R}(D_{s0}^+(2317))$ has been 
postdicted to be $\mathcal{R}(\{D^*K\})\simeq 0.05$ which is much 
smaller than the measured~\cite{PDG06} 
$\mathcal{R}(D_{s1}^+(2460))_{\rm exp}=0.31\pm 0.06$. 
Therefore, all the assignments of $D_{s0}^+(2317)$ to an iso-singlet 
state should be ruled out. 
\begin{center}\hspace{-20mm} 
\begin{figure}[b] 
\includegraphics[width=140mm,clip] 
{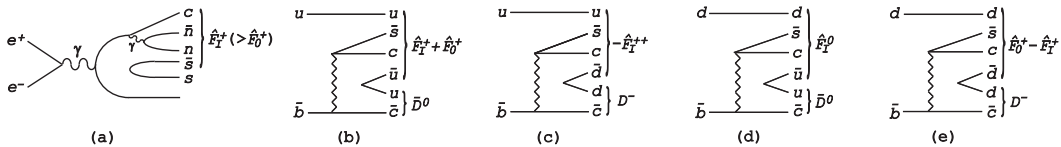}
\caption{
Production of charm strange mesons in $e^+e^-$ annihilation.  
(a) depicts a production of $\hat F_I^+$ (and $\hat F_0^+$).  
(b) and (e) describe productions of $\hat F_I^+$ (and $\hat F_0^+$) 
through $B_u^+$ and $B_d^0$ decays. Productions of $\hat F_I^{++}$ and 
$\hat F_I^0$ through $B_u^+$ and $B_d^0$ decays are described by the 
diagrams (c) and (d), respectively. 
}
\end{figure}\vspace{-5mm}
\end{center} 

As seen above, $D_{s0}^+(2317)$ has been successfully assigned to the  
iso-triplet four-quark meson $\hat F_I^+$ while the other assignments  
have been ruled out. Therefore, its neutral and doubly charged partners, 
$\hat F_I^0$ and $\hat F_I^{++}$, should exist. However, they have not 
yet been observed in inclusive $e^+e^-$ annihilation 
experiments~\cite{Babar-search}. Nevertheless, it does not necessarily 
mean their non-existence, because whether they can be observed or not 
depends on their production mechanism. With this in mind, we study 
productions of $\hat F_I^{++,+,0}$ and $\hat F_0^+$ mesons, and discuss 
why inclusive $e^+e^-$ annihilation 
experiments~\cite{Babar-D_s,CLEO-D_s} have observed $D_{s0}^+(2317)$ 
but did not observe its neutral and doubly charged partners. To study 
productions of scalar $[cn][\bar s\bar n]$ mesons in $e^+e^-$ 
annihilation and $B$ decays, we draw quark-line diagrams within the 
minimal $\{q\bar q\}$ pair creation~\cite{production-B-decay}, because 
multi-$\{q\bar q\}$ pair creation would be suppressed due to the OZI 
rule. The diagram (a) in {FIG.~1} depicts a most probable 
production of $\hat F_I^+$ (and a suppression of $\hat F_0^+$ 
production because of a small $\gamma\{n\bar n\}_{I=0}$ coupling) in 
the $e^+e^-\rightarrow c\bar c$ annihilation, and the diagrams (b) and 
(e) describe their productions through $B_u^+$ and $B_d^0$ decays. If 
the production of charm-strange tetra-quark mesons described by the 
diagram (a) is the main mechanism in the $e^+e^-\rightarrow c\bar c$ 
annihilation, the $\hat F_I^0$ and $\hat F_I^{++}$ production will be 
strongly suppressed, because there is no diagram describing their 
production. On the other hand, $\hat F_I^+$ has been observed in the 
$B_u^+\rightarrow 
\bar D^0({\rm or}\,\,\bar D^{*0})\hat F_I^+$ 
and 
$B_d^0\rightarrow D^-({\rm or}\,\,D^{*-})\hat F_I^+$ 
decays, which are depicted by the diagrams (b) and (e), respectively. 
Productions of $\hat F_I^{++}$ and $\hat F_I^0$ are expected in the 
decay  
$B_u^+\rightarrow D^- ({\rm or}\,\,D^{*-})\hat F_I^{++}$ 
as seen in the diagram (c) and in the decay  
$B_d^0\rightarrow \bar D^0({\rm or}\,\,\bar D^{*0})\hat F_I^0$  
as seen in the diagram (d), respectively, where the diagrams (c) and (d) 
are equivalent to (b) and (e), respectively, under the isospin 
symmetry. Therefore, their production rates are expected to be not very 
far from those of $F_I^{+}$, and hence the branching fractions for 
$\hat F_I^{++}$ and $F_I^{0}$ productions can be estimated 
as~\cite{production-B-decay}  
${Br}(B_u^+\rightarrow D^-\hat F_I^{++})
\sim {Br}(B_u^+\rightarrow \bar D^0
\tilde D_{s0}^+(2317)[D_s^+\pi^0])_{\rm Babar}
= (1.0\pm 0.3\pm 0.1^{+0.4}_{-0.2})\times 10^{-3}$, 
${Br}(B_d^0\rightarrow \bar D^0\hat F_I^{0})
\sim {Br}(B_d^0\rightarrow D^-
\tilde D_{s0}^+(2317)[D_s^+\pi^0])_{\rm Babar} 
= (1.8\pm 0.4\pm 0.3^{+0.6}_{-0.4})\times 10^{-3}$, 
where the last equalities have been taken from Ref.~\cite{Babar-B}.  

In summary, we have studied co-existence of the open-charm conventional 
and tetra-quark scalar mesons. As for the former, the estimated widths 
are $\Gamma(D_{s0}^{*+})\sim 30-40$ MeV and $\Gamma(D_0^*)\sim 40-50$ 
MeV which is much narrower than that of the measured broad $D\pi$ 
enhancement. 

Because the observed scalar nonet mesons, $a_0(980)$, 
$f_0(980)$, $f_0(600)$ and $\kappa(800)$, can be well understood by the 
$[qq][\bar q\bar q]$ mesons, they have been extended to open-charm 
system, and $D_{s0}^+(2317)$ has been successfully assigned to 
$\hat F_I^+$. It has also been discussed that all the members of the 
scalar $[cq][\bar q\bar q]$ mesons are narrow. Therefore, it is awaited 
that experiments re-analyze more precisely the observed enhancement 
($D_0$) just below $D_2^*$ in the $D\pi$ mass distribution, and find 
its structure including $D_0^*$ and $\hat D$. 

Next, it has been demonstrated that the ratio of the decay rates, 
$\mathcal{R}(\hat F_I^+)$, satisfies well the experimental constraint, 
while $\mathcal{R}(\hat F_0^+)$ and $\mathcal{R}(D_{s0}^{*+})$ are  
far beyond the experimental upper bound. In this way, it has been  
concluded that $D_{s0}^+(2317)$ should be assigned to $\hat F_I^+$ but  
its assignments to $D_{s0}^{*+}$ and $\hat F_0^+$ should be ruled out.  
The other existing theories which have postdicted 
$\mathcal{R}(D_{s0}^+(2317))$ are critically reviewed. 
The above discussion implies that $\hat F_I^0$ and $\hat F_I^{++}$ 
should exist. It has been argued that $B$ decays would be better to 
search for $\hat F_I^0$ and $\hat F_I^{++}$ than inclusive 
$e^+e^-\rightarrow c\bar c$ annihilation, and their production rates 
are expected to be 
$Br(B_u^+\rightarrow D^-\hat F_I^{++}) 
\sim Br(B_d^0\rightarrow \bar D^0\hat F_I^{0})\sim 10^{-3}$.   

It is awaited that experiments will confirm co-existence of the 
conventional and tetra-quark mesons in near future. 

\section*{Acknowledgments}
The author would like to thank Professor G.~Rupp, the chair person of 
the workshop, for financial supports. 

\end{document}